\documentclass[aps,showpacs,nofootinbib,12pt]{revtex4}
\usepackage{color}
\usepackage{amsmath}
\usepackage{amsfonts}
\usepackage{amssymb}
\usepackage{mathrsfs}
\newtheorem{definition}{Definition}

\newcommand{\Lam}{{\Lambda}}

\newcommand{\lam}{{\lambda}}

\newcommand{\De}{{\Delta}}

\newcommand{\eps}{{\epsilon}}
\newcommand{\vps}{{\varepsilon}}
\newcommand{\hvps}{{\hat{\varepsilon}}}
\newcommand{\bvps}{{\bar\varepsilon}}
\newcommand{\hbvps}{{\hat{\bar\varepsilon}}}

\newcommand{\al}{\alpha}
\newcommand{\bal}{{\bar\alpha}}
\newcommand{\be}{\beta}
\newcommand{\bbe}{{\bar\beta}}
\newcommand{\de}{\delta}
\newcommand{\bde}{{\bar\delta}}
\newcommand{\bxi}{{\bar\xi}}
\newcommand{\bX}{{\bar X}}
\newcommand{\del}{{\nabla}}

\newcommand{\bm}{{\bar m}}

\newcommand{\bmu}{{\bar\mu}}

\newcommand{\bpi}{{\bar\pi}}

\newcommand{\brho}{{\bar\rho}}
\newcommand{\bome}{{\bar\omega}}

\newcommand{\bze}{{\bar\zeta}}

\def\d#1#2{\displaystyle\frac{\displaystyle #1}{\displaystyle #2}}
\newcommand{\ppt}{{{\partial_t}}}
\newcommand{\ppr}{{{\partial_r}}}
\newcommand{\ppu}{{{\partial_u}}}
\newcommand{\ppz}{{{\partial_\zeta}}}
\newcommand{\ppbz}{{{\partial_{\bar\zeta}}}}
\newcommand{\pppt}{{\d{\partial}{\partial t}}}
\newcommand{\ppR}{{{\partial_R}}}
\newcommand{\heq}{{\ {\hat =}\ }}
\newcommand{\cL}{{\cal L}}
\newcommand{\cH}{{\cal H}}

\newcommand{\cA}{{\cal A}}
\newcommand{\cB}{{\cal B}}

\def\r{{\partial}}

\newcommand{\omits}[1]{}
\definecolor{dyellow}{rgb}{1.,0.8,.0}
\definecolor{myblue}{rgb}{.1,.1,.7}
\definecolor{dcyan}{rgb}{.0,.6,.6}
\definecolor{dmagenta}{rgb}{0.6,0.0,0.6}
\definecolor{brown}{rgb}{0.6,0.2,0.}
\definecolor{darkblue}{rgb}{.0,.0,0.5}
\definecolor{darkred}{rgb}{0.75,0.0,0.0}
\definecolor{orange}{rgb}{1.,.6,.0}
\definecolor{dorange}{rgb}{0.8,.4,.0}
\definecolor{darkgreen}{rgb}{0.0,0.6,0.0}
\definecolor{purple}{rgb}{.4,.0,.4}

\def\red{\color{red}}

\begin{document}
\title{On Gravitational anomaly and Hawking radiation near weakly isolated horizon}
\author{Xiaoning Wu$^{a,c}$\footnote{Email: wuxn@amss.ac.cn},
Chao-Guang Huang$^{b,c}$\footnote{Email: huangcg@ihep.ac.cn}, and
Jia-Rui Sun$^{b,d}$\footnote{Email: sun@ihep.ac.cn}}
\affiliation{\footnotesize $^a$Institute of Mathematics, Academy of
Mathematics and Systems Science, Chinese Academy of Sciences,
P.O.Box 2734,
Beijing, 100080, China} %
\affiliation{\footnotesize
$^b$Institute of High Energy Physics, Chinese Academy of Sciences,
P.O. Box 918(4), Beijing, 100049, China}%
\affiliation{\footnotesize  $^c$Kavli Institute for
Theoretical Physics China at the Chinese Academy of Sciences
(KITPC-CAS), P.O.Box 2732,
Beijing, 100080, China} %
\affiliation{\footnotesize $^d$Graduate School of Chinese Academy
of Sciences, Beijing, 100049, China.}


\bigskip

\begin{abstract}
Based on the idea of the work by Wilczek and his collaborators, we
consider the gravitational anomaly near weekly isolated horizon.
We find that there exists a universal choice of tortoise
coordinate for any weakly isolated horizon.  Under this
coordinate, the leading behavior of a quite arbitrary
scalar field near horizon is a 2-dimensional chiral scalar field.
This means we can extend the idea of Wilczek and his collaborators to
more general cases and show the relation between
gravitational anomaly and Hawking radiation is a universal
property of black hole horizon.

\end{abstract}

\pacs{04.70.Dy} 

\maketitle

\newpage

\section{Introduction}

Hawking radiation and black hole thermal dynamics is believed to be
most important evidence for the deep relation between quantum theory
and general relativity. After Hawking's original
paper\cite{Ha74,Ha75}, many works have been done in this area in
order to get deeper understanding on the properties of quantum
fields in curved spacetimes. Many methods have been developed in
last several decades. They all show that the phenomena of Hawking
radiation exists for many kinds of black hole
\cite{Un74}-\cite{MV05}, including for some non-stationary black
holes \cite{Ba86}-\cite{ZZZ94}. Recently, Wilczek and his
collaborators proposed a new method based on the anomaly analysis
\cite{RW05,IUW06}. The anomaly analysis in the study of the Hawking
radiation can be traced back to Christensen and Fulling \cite{CF77}.
They consider the trace anomaly of a conformally invariant scalar
field in Schwarzschild background and show that there is a relation
between the Hawking radiation and anomalous trace of the field under
the condition that the covariant conservation law is valid. By
imposing a boundary condition near horizon, Wilczek and his
collaborators prove that the Hawking radiation is just the cancel
term of the gravitational anomaly of the covariant conservation law
\cite{RW05} and gauge invariance as well\cite{IUW06}.  Later,  this
idea is extended to other kinds of black
holes\cite{MS06}-\cite{HSW}. The aim of this paper is to generalize
the approach of Wilczek and his collaborators to more general cases,
including dynamical black holes.   To do so,  a general definition
of a horizon is needed since the event horizon of black hole cannot
describe the dynamical properties of black hole very well
\cite{As04}. The weakly isolated horizon is the appropriate notation
to replace the event horizon.  Thus, we shall show in this paper
that the approach is available to the weakly isolated horizons.

The organization of this paper is as follows.  In section II, we
briefly review the definition of weakly isolated horizon
and the geometry near
it
in the Bondi-like coordinate system with Bondi-gauge. Some
detail analysis on the asymptotic behavior of the d'Alembert
operator near horizon is also made in this section. Section III
gives the calculation of gravitational anomaly near a weakly
isolated horizon. Section IV focuses on the gauge anomaly of
electromagnetic field. In section V, the analysis is generalized
to weakly isolated horizons in higher dimensional spacetimes.
Section VI contains some discussions.


\section{Preliminaries}

During last decade, motivated by the need of numerical relativity
and relativistic astrophysics, much work has been done to find a
quasi-local definition of a black hole. A very nice review on the
development on the quasi-local definition of a horizon and its
possible applications can be found in Ref.\cite{As04}. In the
present paper, we will follow the definition of the weakly isolated
horizon, given by Ashtekar and his collaborators.%
\begin{definition}(Weakly Isolated Horizon)\\
Let $(M, g)$ be a space-time . $\cH$ is a 3-dim null hyper-surface
in $M$ and $l^a$ is the tangent vector field of the generator of
$\cH$. $\cH$ is said to be a {\bf weakly isolated horizon}(WIH),
if\\
1) $\cH$ has the topology of $S^2\times{\bf R}$;\\
2) The expansion of the null generator of $\cH$ is zero, i.e.
$\Theta_l=0$ on $\cH$;\\
3) $T_{ab}v^b$ is future causal for any future causal vector
$v^a$ and Einstein equation holds in a neighborhood of $\cH$;\\
4)
$[\cL_l\ ,\ D_a]l^b=0$ on $\cH$, where $D_a$ is the induced
covariant derivative on $\cH$.
\end{definition}%
By definition, it can be shown that there exists a 1-form
$\omega_a$
on $\cH$ such that $D_al^b\heq\omega_al^b$, where $\heq$ denotes the
equality restricted to $\cH$.  Similarly, for quantity $f$ on
spacetime, ${\hat f}$ means the value of $f$ on horizon $\cH$,
following the notation in Ref.\cite{As04}.

As in Ref.\cite{WG07}, we can introduce a Bondi-like coordinates
$(u,r,\theta,\varphi)$ in a neighborhood of horizon $\cH$ in the
following way.  First, denote the tangent vector of null generator
of $\cH$ as $l^a$ and another real null vector field as $n^a$. Then,
the foliation of $\cH$ gives us the natural coordinates $(\theta,
\varphi)$. Their Lie drag along each generator of $\cH$ together
with the parameter $u$ of $l^a$ form the coordinates on $\cH$.
Finally, choose the affine parameter $r$ of $n^a$ as the forth
coordinate.  Furthermore, we can also choose a set of null tetrad
which satisfy Bondi gauge in this neighborhood \cite{PR86, NT80}.
The expression of
the tetrad in Bondi coordinates are 
\begin{eqnarray}
\begin{cases}
l^a = {\partial_u}+U{\partial_
r}+X{\partial_\zeta}+\bX{\partial_\bze}\ , & \cr
n^a = {\partial_r}\ , &\cr
m^a=\omega{\partial_r}+\xi^3\ppz+\xi^4\ppbz\ ,& \cr
\bm^a=\bome\ppr+\bxi^4\ppz+\bxi^3\ppbz\ ,
\end{cases}\label{tetrad}
\end{eqnarray}
where%
\omits{\begin{eqnarray}
U= O(r),  \qquad  X=O(r^2),  \qquad
\omega = O(r^2), \qquad \xi^A = O(1),
\end{eqnarray}}
\begin{eqnarray}
U \heq X \heq \omega \heq 0, \label{UXom}
\end{eqnarray}
and $\zeta=e^{i\varphi}\cot\frac{\theta}{2} $.
\omits{on $\cH$ (following the notation in
ref.\cite{As04}, equalities restricted to $\cH$ will be denoted by
`` $\heq$ ". Otherwise, for quantity $f$ on spacetime , ${\hat f}$
means the value of $f$ on horizon $\cH$),}
Then the metric takes the
form \cite{NT80}
\begin{eqnarray}
\left(g^{\mu\nu}\right)=\left(\begin{array}{cccc}
0&1&0&0\\
1&2(U-|\omega|^2)&X-(\bome\xi^3+\omega\bxi^3)&\bX-(\bome\xi^4+\omega\bxi^4)\\
0&X-(\bome\xi^3+\omega\bxi^3)&-2|\xi^3|^2 &-(\xi^3\bxi^4+\bxi^3\xi^4)\\
0&\bX-(\bome\xi^4+\omega\bxi^4)&-(\xi^3\bxi^4+\bxi^3\xi^4)
&-2|\xi^4|^2 \end{array}\right)\label{metric}.
\end{eqnarray}
It is easy to see that $\sqrt{-g}=\sqrt{h}$ if we denote
$(h_{AB}):=-(\xi^A\bxi^B+\bxi^A\xi^B)^{-1}$ with $A, B= 3,4$.
Obviously, $h_{AB}$ is the induced metric on the section of WIH.

In the Newman-Penrose formalism, the Bondi gauge can be
expressed as
\begin{eqnarray}
\nu=\tau=\gamma=\al+\bbe-\pi=\mu-\bmu=0, \qquad 
\vps-\bvps\ \heq\ \kappa\ \heq\ 0,
\label{gaugec}
\end{eqnarray}
where
\begin{eqnarray*}
-\nu &=&  n_{\mu;\nu}\bm^\mu n^\nu = \bm^\mu \Delta n_\mu, \\
\tau &=& l_{\mu;\nu} m^\mu n^\nu =m^\mu \Delta l_\mu, \\
-\pi &=&  n_{\mu;\nu}\bm^\mu l^\nu = \bm^\mu D n_\mu,\\
-\mu &=&  n_{\mu;\nu}\bm^\mu m^\nu = \bm^\mu \delta n_\mu,\\
\kappa &=& l_{\mu;\nu}m^\mu l^\nu = m^\mu D l_\mu,\\
- \gamma &=& \frac 1 2 (n_{\mu;\nu}l^\mu n^\nu-\bm_{\mu;\nu} m^\mu
n^\nu) =\frac 1 2 (l^\mu \Delta n_\mu
- m^\mu \Delta  \bm_\mu) , \\
- \al &=& \frac 1 2 (n_{\mu;\nu}l^\mu \bm^\nu-\bm_{\mu;\nu} m^\mu
\bm^\nu) =
\frac 1 2 (l^\mu \bar \delta n_\mu - m^\mu \bar \delta \bm_\mu)  , \\
\be &=&  \frac 1 2 (l_{\mu;\nu}n^\mu m^\nu-m_{\mu;\nu} \bm^\mu
m^\nu)
=\frac 1 2 (n^\mu  \delta l_\mu - \bm^\mu  \delta m_\mu), \\
\vps &=& \frac 1 2 (l_{\mu;\nu}n^\mu l^\nu - m_{\mu;\nu} \bm^\mu
l^\nu)= \frac 1 2 (n^\mu  D l_\mu - \bm^\mu  D m_\mu) ,
\end{eqnarray*}
where $D:=l^a \del_a$, $\Delta :=n^a \del_a$, $\de:= m^a \del_a$ and
$\bde:=\bm^a \del_a$ as the standard notation in \cite{PR86}. Other
three spin coefficients are
\begin{eqnarray*}
\rho &=& l_{\mu;\nu} m^\mu \bm^\nu = m^\mu \bde l_\mu, \\
\sigma &=& l_{\mu;\nu} m^\mu m^\nu = m^\mu \de l_\mu, \\
-\lambda &=&  n_{\mu;\nu} \bm^\mu \bm^\nu  = \bm^\mu\bde n_\mu.
\end{eqnarray*}
In the Bondi gauge, 1-form $\omega_a$ is expressed as
\[
\omega_a=-(\vps+\bvps)n_a+(\al+\bbe)\bm_a+(\bal+\be)m_a =
-(\vps+\bvps)n_a+\pi \bm_a+\bar \pi m_a ,
\]
and $(\vps+\bvps)|_{\cH}$ is constant \cite{As04}.  In addition, the
definition of WIH implies
\begin{eqnarray}
\rho \heq \sigma \heq 0. \label{rhosi}
\end{eqnarray}
Based on the result in Ref.\cite{As04}, the angular momentum of WIH
is
\[
J[\varphi]=-\frac{1}{8\pi}\int_S(\varphi^a\omega_a)dV_2,
\]
where $\varphi^a$ is a vector field on section $S$.  So, among
the NP coefficients, only
$\pi$ is related to the angular momentum of WIH.

The commutators of the null tetrad require
\begin{eqnarray}
\begin{array}{llll}
U= (\hvps+\hbvps)r+O(r^2), &  \kappa = O(r), &  \al = O(1), & \mu=O(1), \smallskip\\
\vps = \hvps+O(r), & \d {\partial \xi^3}{\partial u} \heq 0, & \be = O(1), &\lam = O(1), \smallskip \\
\pi = O(1), & \d {\partial \xi^4}{\partial u} \heq 0, &  &
\omega = O(r).
\end{array} \label{str}
\end{eqnarray}

The d'Alembert operator can be re-written as
\begin{eqnarray}
\Box&=&
(l^an^b+n^al^b-m^a\bm^b-\bm^am^b)\del_a\del_b\nonumber\\
&=&D\Delta +\Delta D +(\mu+\bmu)D+(\vps+\bvps)\Delta - (\rho+\brho)\Delta -\pi\de-\bpi\bde-\Delta_S\nonumber\\
&=&2\ppu\ppr+2U\left(\ppr\right)^2+2X\ppr\ppz+2\bX\ppr\ppz\nonumber\\
&&+\frac{\partial U}{\partial r}\ppr+\frac{\partial X}{\partial
r}\ppz+\frac{\partial\bX}{\partial r}\ppbz + (\mu+\bmu)\left (\ppu+\omits{(\mu+\bmu)}U\ppr+\omits{(\mu+\bmu)}X\ppz
+\omits{(\mu+\bmu)}\bX\ppbz\right )\nonumber\\
&&+(\vps+\bvps)\ppr -
(\rho+\brho)\ppr-\pi\de-\bpi\bde-\de\bde-\bde\de +(\al - \bar\be)\de
+(\bar \al -\be)\bde.\label{box}
\end{eqnarray}
Detail calculation also tells us
\[
\De_S = \de\bde+\bde\de - (\al - \bar\be)\de - (\bar \al
-\be)\bde+O(r^2)
\]
where $\De_S$ is the Laplacian on the coordinate 2-sphere.


\section{Gravitational anomaly \label{Sec:anomaly}}
In this section, we will show that the relation between Hawking
radiation and gravitational anomaly still exists for weakly isolated
horizons.

Let's consider a scalar field near a weakly isolated horizon, whose
action can be written as
\begin{eqnarray}
S = - \frac 1 2 \int dV_4\ \phi(\Box-m^2)\phi.\label{action}
\end{eqnarray}
Using Eq.(\ref{box}), it can be written in the explicit form in
Bondi-like coordinate system
\begin{eqnarray}
S&=&-\frac 1 2 \int du dr d\theta
d\varphi\sqrt{h}\ \phi\left[2\ppu\ppr+2U\left(\ppr\right)^2 + 2X\ppr\ppz+2\bX\ppr\ppz\right.\nonumber\\
&&+(\partial_r U) \ppr+(\partial_r X)\ppz+(\partial_r\bX)\ppbz+(\mu+\bmu) \left ( \ppu+ U\ppr + X\ppz
+ \bX \ppbz \right )\nonumber\\
&& + (\vps+\bvps) \ppr -(\rho+\brho)\ppr -\pi (\omega {\partial_r}
+ \xi^3 {\partial_\zeta} + \xi^4 {\partial_{\bar \zeta}})\nonumber\\
&&\left.-\bpi (\bar \omega {\partial_r} + \bar \xi^4
{\partial_\zeta} + \bar \xi^3 {\partial_{\bar \zeta}})-
\Delta_S+O(r^2)-m^2\right]\phi  \label{m1}
\end{eqnarray}

Introduce new coordinates in a neighborhood of WIH as
\begin{eqnarray}
t:=u-r_*, \qquad R:=r,\label{tR}
\end{eqnarray}
where $dr_*=dr/f(r)$ and $f(r)=2(\hvps+\hbvps)r+O(r^2)$.
Then it is easy to get
\begin{eqnarray}
{\partial_u}={\partial_t}\ ,\qquad
{\partial_r}=- f^{-1}{\partial_t}+{\partial_R}\ .\label{ppur}
\end{eqnarray}
Denote $\{\lam_k\}$ are eigenvalues of the Laplacian $\De_S$ and
$\{F_k\}$ are associated (normalized) eigenfunctions, and make the
variable separation for the scalar field $\phi$ as
\begin{eqnarray}
\phi(u,r,\theta,\varphi)=\sum_k\phi_k(u,r)F_k(\theta,\varphi).\label{sep}
\end{eqnarray}
Then, the action becomes
\begin{eqnarray}
S&=&-\frac 1 2 \sum_k\int dtdR\cdot\phi_k\left[
-\frac{1+O(r)}{f}\left(\ppt\right)^2 +[1+O(r)]\ppR (f \ppR)
+O(r)\ppt\ppR \right.\nonumber\\
&&+\frac{O(r)}{f}\ppt+\frac 1 2[\mu+\bmu +O(r)]\ppt+(\mu+\bmu)[(\hvps+\hbvps)r+O(r^2)]\ppR \nonumber\\
&&\left.+\frac{\rho+\brho}{f}\ppt-(\rho+\brho)\ppR+ \frac {\pi \omega + \bpi \bar \omega}{f}\ppt - (\pi \omega
+ \bpi \bar \omega)\ppR -\lam_{k}+\tilde{O}(r)\right] \phi_k\nonumber\\
&&-\frac 1 2 \sum_{k,k'} \int dt dR  f^{-1}\phi_k (t,R) O_{kk'}(t,R)
\phi_{k'} (t,R) .
\end{eqnarray}
Here the symbol $\tilde{O}(r)$ represents an $O(r)$ operator without
$O(r)\ppr$ terms and
\begin{eqnarray}
O_{kk'}(t,R)&=&\int d\theta d\varphi \sqrt{h} \ F_k(\theta,\varphi)
\left [ 2X\left (-\ppt + f \ppR\right )\ppz+2\bX \left (- \ppt + f\ppR \right )\ppz\right .\nonumber\\
&& +\left (-{\partial_t X} + f {\partial_R X}\right ) \ppz
+\left (-{\partial_t \bX}  + f{\partial_R \bX}\right )\ppbz
+(\mu+\bmu) f\left (X\ppz + \bX \ppbz \right )\nonumber\\
&&\left .-\pi f \left (\xi^3 {\partial_\zeta} + \xi^4
 {\partial_{\bar \zeta}}\right ) -\bpi f\left (\bar
\xi^4 {\partial_\zeta} + \bar \xi^3
{\partial_{\bar \zeta}}\right ) \right]F_{k'}(\theta,\varphi)
\end{eqnarray}
is another $\tilde{O}(r)$ operator.  In the tortoise coordinate
$dR_*=dR/f$, the action reduces to
\begin{eqnarray}
\omits{&&-\frac 1 2\int dV^4\ \phi(\Box+m^2)\phi\nonumber\\
&=&-\frac 1 2\sum_k\int dt dR_* \  f\phi_k\left[-\frac{1+O(r) } {f}
\left(\ppt\right)^2 +[1+O(r)]\ppR f\ppR +O(r)\ppt\ppR
\right.\nonumber\\
&& +\frac{O(r)}{f}\pppt
+\left (\frac{\rho+\brho}{f} +\frac{1}{2}(\mu+\bmu)[1+O(r)]
\right)\pppt+ (\mu+\bmu)r[\hvps+\hbvps+O(r)]\ppR\nonumber\\
&&\left. -(\rho+\brho)\ppR  - (\pi \omega + \bpi \bar \omega)\ppR-\lam_k+m^2+\tilde{O}(r)\right]\phi_k \nonumber\\
&& -\frac 1 2 \sum_{k,k'} \int dt dR_* \phi_k (t,R) O_{kk'}(t,R) \phi_{k'} (t,R)\nonumber\\}
S&=&-\frac 1 2\sum_k\int dtdR_*\
\phi_k\left[-\left(\ppt\right)^2+\left({\partial_{
R_*}}\right)^2+O(r)\right]\phi_k\nonumber\\
&& -\frac 1 2 \sum_{k,k'} \int dt dR_* \phi_k (t,R) O_{kk'}(t,R)
\phi_{k'} (t,R).
\end{eqnarray}
With respect to coordinate $R_*$, the term $O(r)$ vanishes
exponentially, so the dominant term of the action (\ref{action})
near a  weakly isolated horizon takes the form of an infinite
collection of 2-dimensional fields. The metric of the effective
2-dimensional spacetime  is
\begin{eqnarray}
ds^2=-f(R)dt^2+\frac{dR^2}{f(R)},\label{2metric}
\end{eqnarray}
and the horizon is the boundary of the effective spacetime. This
behavior is similar to what happens near Schwarzschild black hole
horizon \cite{RW05}.

As in Ref. \cite{RW05, MS06}, the effective 2-dimensional spacetime
is bounded by the horizon on one side (at $R=r=0$), on which the
boundary condition that the outgoing modes vanish is imposed.  Then,
in the ``near horizon region", $0<r<a$ with $a\to 0$, the fields
become chiral.  It is well-known that a $(4k+2)$-dimensional chiral
theory contains the following gravitational
anomaly\cite{Be00}-\cite{BK01}
\begin{eqnarray}
\del_{\mu}T^{\mu}_{\nu}\omits{\red =A_{\nu}=\frac{1}{\sqrt{-g}}\
\partial_{\mu}N^{\mu}_{\nu}}=\frac{1}{96\pi\sqrt{-g}}\
\eps^{\be\al}\partial_{\al}\partial_{\eta}\Gamma^{\eta}_{\nu\be},
 \label{anomaly0}
\end{eqnarray}
where $\al,\be,\eta,\mu,\nu=0,1$, $g$ is the determinant of
$(4k+2)$-dimensional metric, and the convention $\eps^{01}=1$ is
used. The divergence of the energy-momentum tensor can be generally
written as
\begin{eqnarray}
\del_{\mu}T^{\mu}_{\nu}={\mathscr A}_{\nu}=\frac{1}{\sqrt{-g}}\
\partial_{\mu}N^{\mu}_{\nu} .\label{anomaly}
\end{eqnarray}
In the ``out region", $r>a$, there is no anomaly in the divergence
of the energy-momentum tensor. Therefore, $N^{\mu}_{\nu}=0$ and thus
${\mathscr A}_{\nu}=0$. In near horizon region, from the metric
(\ref{2metric}) and Eq.(\ref{anomaly0}), it is easy to see
\begin{eqnarray}
N^t_t=N^R_R=0,\qquad N^R_t=\frac{1}{192\pi}(ff\,')',\qquad
N^t_R=\frac{1}{192\pi }(f^{-1}f\,')',
\end{eqnarray}
where a prime means $\partial_R$ and thus
\begin{eqnarray}
{\mathscr A}_t=\frac{1}{192\pi}(ff')'',\qquad {\mathscr A}_R=0.
\end{eqnarray}

The effective action for the metric after integrating out the field $\phi$ is
\begin{eqnarray}
W[g_{\mu\nu}]=-i\ln\left(\int{\cal
D}[\phi]\exp(iS[\phi,g_{\mu\nu}])\right),
\end{eqnarray}
where $S[\phi,g_{\mu\nu}]$ is the classical action. A basic
requirement for a well-defined quantum theory is that it should be
anomaly free. In the present case, it is equivalent to require that
the full quantum theory is diffeomorphism invariant. The requirement
can be expressed in terms of $W[g_{\mu\nu}]$. Suppose $v^a$ to be a
vector field, $\de_v$ is the variation induced by $v^a$. Under the
variation induced by any $v^a \in TM$,
\begin{eqnarray}
-\de_vW &=& \int d^2x\sqrt{-g}\ v^{\nu}\del_{\mu}[T^{\ \mu}_{\chi\
\nu}\Theta_-+T^{\ \mu}_{O\ \nu}\Theta_+], \nonumber \\
&=&\int d^2x\ v^t[\partial_{R}(N^R_t\Theta_-)+(T^{\ R}_{O\ t}-T^{\
R}_{\chi
\ t}+N^R_t)\partial _R\Theta_+]\nonumber\\
&&+\int d^2x\ v^R(T^{\ R}_{O\ R}-T^{\ R}_{\chi \
R})\partial_R\Theta_+.
\end{eqnarray}
where $T^{\ \mu}_{\chi\ \nu}$ and $T^{\ \mu}_{O\ \nu}$ are the
energy-momentum tensor in near horizon and in the out region,
respectively, as the notation in \cite{RW05}, and
$\Theta_+=\Theta(r-a)$ and $\Theta_-=1-\Theta_+$ are step functions.

To obtain the concrete expression for the variation of effective
action, the explicit expression for the energy-momentum tensor is
needed.  Because the  effective metric is static, $T^{\mu}_{\ \nu}$
is independent of $t$.  Then the general solutions of
Eq.(\ref{anomaly}) for $T^{\ \mu}_{\chi\ \nu}$ and $T^{\ \mu}_{O\
\nu}$ are
\begin{eqnarray}
\begin{cases}
T^t_{\ t}=-\d{K+Q}{f}-\d{B(R)}{f}-\d{I(R)}{f}+T(R)\ , \smallskip \cr
T^R_{\ R}=\d{K+Q}{f}+\d{B(R)}{f}+\d{I(R)}{f}\ , \smallskip \cr
T^R_{\ t}=-K+C(R)=-f^2T^t_R\ ,
\end{cases}\label{solution}
\end{eqnarray}
where
\begin{eqnarray}
\begin{cases}
C(R)=\displaystyle\int^R_0{\mathscr A}_t(s)ds\ ,\smallskip\cr
B(R)=\displaystyle\int^R_0 f(s){\mathscr A}_R(s)ds\ ,\smallskip\cr
I(R)=\d 1 2\displaystyle\int^R_0 T(s)f'(s)ds \, ,
\end{cases}
\end{eqnarray}
and $T(R)$ is the trace of energy-momentum tensor. Therefore, the
variation of the effective action becomes
\begin{eqnarray}
-\omits{\lim_{a\to 0}} \de_vW &=&\omits{\lim_{a\to 0}\left\{}\int d^2x\
v^t[\partial_R(N^R_t\Theta_-)+(N^R_t+K_{\chi}-K_O)\de(R-a)]\nonumber\\
&&\qquad\omits{\left.}+\int d^2x\
v^R\frac{K_O+Q_O-K_{\chi}-Q_{\chi}}{f}\de(R-a) .
\end{eqnarray}
The requirement that $\lim_{a\to 0} \de_v W=0$ for any vector field
$v^a$ demands
\begin{eqnarray}
&&K_O=K_{\chi}+ \Phi \ ,\nonumber\\
&&Q_O=Q_{\chi}- \Phi\ ,\nonumber\\
&&\Phi =N^R_t(0)=\frac{1}{48\pi}(\hvps+\hbvps)^2.
\end{eqnarray}
On the other hand, the surface gravity of a WIH is
${\kappa_l}=l^a\omega_a=(\hvps+\hbvps)$ \cite{As04}. So,
\begin{eqnarray}
\Phi = \frac{\kappa_l^2}{48\pi}
\end{eqnarray}
This means the gravitational anomaly near WIH has similar behavior
as in Schwarzschild spacetime\cite{RW05}.

The total energy-momentum tensor
$T^{\mu}_{~\nu}=T^{\mu}_{\chi\nu}+T^{\mu}_{O\nu}$ can be rewritten,
in the limit $a\to 0$, in two parts:
\begin{eqnarray}
T^{\mu}_{~\nu}=T^{\mu}_{c\nu}+T^{\mu}_{\Phi\nu},\label{ten}
\end{eqnarray}
where $T^{\mu}_{c\,\nu}$ is the conserved energy-momentum tensor of
matter field which behaves as without any quantum effects, and
$T^{\mu}_{\Phi\,\nu}$ is a conserved tensor with $K=-Q={\Phi}$, a
pure flux, which appears as the requirement to cancel the
gravitational anomaly. Since Eq.(\ref{ten}) has the same form as the
flux of black body radiation in $R$ direction at temperature $T$ in
2D spacetime, it is just a thermal radiation with the Hawking
temperature $T=\kappa_l/(2\pi)$.

A remark on the relation between our results and the Planck
distribution is needed. It is well-known that a quantum field theory
in curved space-time deeply depends on the choice of the observer.
Unruh effect is a quite nice example. In this section, the observers
which we used are rest ones in the ``rest frame" in terminology of
Ashtekar {\it et al} \cite{As04}. Concretely, the time direction is
${\partial_t}$ and the coordinate system is $(t,R,\theta,\varphi)$.
In the coordinate system, a mode state of scaler field $\phi(x)$
labeled by quantum number $E$ and $m$ is
\[
\phi\propto\exp(-iEt+im\varphi),
\]
and the distribution function observed by the observer,
following the argument in Ref.\cite{RW05,IUW06,IUW062}, is $(\exp(E/T)+1)^{-1}$.

However, as emphasized
in Ref.\cite{As04}, not every choice of time direction will result in a
Hamiltonian evolution in the phase space, then a horizon mass and
first law of black hole thermal dynamics. 
In order to obtain the black hole mechanical law, one has to choose
the canonical time. In non-rotational cases, the difference between
our $t$ and the canonical one is higher order terms so it makes no
contribution. In contrast, the non-zero horizon angular momentum
will change all things. In the latter case, the leading term of the
canonical time is
\begin{eqnarray}
\partial_{t_c}\heq \partial_t + \Omega_t \partial_\varphi ,
\end{eqnarray}
where $\partial_\varphi$ is the Killing vector for the metric on the
2-dimensional section of horizon and $\Omega_t$ is the angular
velocity of the horizon. What is interested in is the radiation seen
by the canonical observers. \omits{To do so, we introduce a new
coordinates $(t_c, R_c, \theta_c, \varphi_c)$ as
\begin{eqnarray}
t_c=t,\ R_c=R,\ \theta_c=\theta,\ \varphi_c=\varphi-\Omega_tt.
\end{eqnarray}}
In the
coordinates of a canonical observer near the horizon, $(t_c, R_c, \theta_c, \varphi_c)$, which is defined by
\begin{eqnarray}
t_c=t,\ R_c=R,\ \theta_c=\theta,\ \varphi_c=\varphi-\Omega_tt,
\end{eqnarray}
the mode state
should be
\[
\phi\propto\exp[-i(E-m\Omega_t)t_c+im\varphi_c].
\]
Then
the distribution function observed by a canonical observer should be
\[
\frac{1}{\exp(\frac{E-m\Omega_t}{T})+1}\ . \qquad
\]
This is the Planck spectrum with non-zero chemical potential. It
means that a rotational isolated horizon has the same radiation spectrum as
Kerr black hole \cite{MS06,IUW062}.


\section{Gauge Anomaly}
\omits{In last section, we considered the gravitational anomaly near
weakly isolated horizon.} Now, let us turn to consider the gauge
anomaly near WIH. The action of a complex scalar field $\phi(x)$
near WIH coupled to electromagnetic field is
\begin{eqnarray}
S&=&\frac 1 2 \int dV_{4} \, \{
[(\del_a-ieA_a){\bar\phi}]\cdot[(\del^a+ieA^a)\phi]+m^2|\phi|^2\}\nonumber\\
&=&{\frac 1 2 }\int dV_{4}\, \{ {\bar\phi}[-\Box +m^2]\phi -
ie{\bar\phi}A_a\del^a\phi + ie\phi A_a\del^a{\bar\phi} +
e^2{\bar\phi}|A|^2\phi\}.
\end{eqnarray}
In the second equality a surface term is omitted.

The first interaction term in the integrand reads in the coordinate
systems used in the previous sections
\begin{eqnarray}
-ie \bar \phi A_a\del^a\phi &=& -ie \bar \phi \left[f^{-1}\left(-{A_u} +fA_r-\cA
A_r-A_A\cB^A\right)\ppt\right.\nonumber\\
&&\quad\left.+\left(A_u+\cA
A_r+A_A\cB^A\right)\ppR+\left(A_r\cB^A+h^{AB}A_B\right)\partial_A\right]\phi\nonumber\\
&=&-ie f^{-1}\bar \phi \left[-A_u\ppt+\left(A_u+\cA
A_r+A_A\cB^A\right){\partial_{R_*}}+\tilde{O}(r)\right]\phi\ ,
\end{eqnarray}
where $\cA=2(U-|\omega|^2)$, $\cB^3=X-(\bome\xi^3+\omega\bxi^3)$,
$\cB^4=\bX-(\omega\bxi^3+\bome\xi^3)$, $\partial_3= \partial_\zeta$,
$\partial_4 = \partial_{\bar \zeta}$. Under the coordinate
transformation (\ref{ppur}),
\begin{eqnarray}
A_t=A_u,\ A_R=f^{-1}A_u+A_r,\ A_{\zeta}=A_{\zeta}\ \mbox{ and }\
A_{\bar \zeta}=A_{\bar \zeta}\, . \label{AtAu}
\end{eqnarray}
Thus,
\begin{eqnarray}
-i e\bar \phi A_a\del^a\phi = -i ef^{-1}\bar \phi \left[-A_t\ppt+\left(f
A_R+O(r)\right){\partial_{R_*}}+\tilde{O}(r)\right]\phi \, .
\end{eqnarray}
Suppose $A_a$ be in the Ashtekar gauge, defined by \cite{As04,As00}
\begin{eqnarray}
\cL_lA_{\underleftarrow{a}} \heq 0 , \label{Aga}
\end{eqnarray}
{where an arrow ``$\underleftarrow{\ }$" denotes the pullback to
$\cH$, e.g. $A_{\underleftarrow{a}}$ denotes the pullback of $A_a$
to $\cH$. In the Ashtekar gauge,} $\Phi_l:=-l^aA_a$ is constant on
$\cH$, which is the analogue of the static electric potential on the
Reissner-Nordstr\"om horizon \cite{IUW06}. The potential
$A'_a=A_a+(d\al)_a$ is also in the Ashtekar gauge if and only if
\begin{eqnarray}
l\cdot d\al = D\al =\partial_u\al +U\partial_r \al + X\partial_\zeta+\bar X\partial_{\bar \zeta} \heq \partial_u\al
+U\partial_r \al\heq C. \label{Dalpha}
\end{eqnarray}
Therefore, for a given $A_a$ satisfying (\ref{Aga}) with a nonzero
$A_R$ on $\cH$,  there exists a gauge transformation satisfying
(\ref{Dalpha}) and
\begin{eqnarray}
\partial_r \al = -(2A_R+U^{-1}C),
\end{eqnarray}
which makes $A'_a$ satisfy (\ref{Aga}) with
\omits{insures the Ashtekar gauge and}
\[A'_R =f^{-1}A'_u+A'_r=f^{-1}(A_u+\partial_u\al)+A_r+\partial_r\al
\heq A_R +\frac 1 2 \partial_r\al + f^{-1}C\heq 0.\]  Namely, one may always write
\begin{eqnarray}
-ie\bar \phi A_a \nabla^a \phi = ief^{-1}\bar \phi \left[A_t\ppt+\tilde{O}(r)\right]\phi \,
\end{eqnarray}
without loss of generality.  The second interaction term is just the
complex conjugate of the first one.  The last interaction term reads
in the above gauge
\begin{eqnarray}
e^2|A|^2 |\phi|^2= e^2(2A_uA_r+\cA A^2_r+2A_rA_B\cB^B+h^{AB}A_AA_B)|\phi|^2 \heq
-e^2f^{-1}A_u^2|\phi|^2.
\end{eqnarray}
Therefore, under above gauge choice, the interaction part of the
action in near horizon region can be written as
\begin{eqnarray}
\sum_k\int dtdR_*\left\{ie{\bar\phi_k}\left(A_t\pppt\right)\phi_k
-ie{\phi_k}\left(A_t\pppt\right){\bar\phi}-e^2|\phi_k|^2{A^2_t}+O(r)\right\}
\end{eqnarray}
and the action of the complex scalar field as
\begin{eqnarray}
S=\sum_k\int dtdR_*\
 {\bar\phi_k}\left[-\left(\pppt-ieA_t\right)\left(\pppt+ieA_t\right)
 +\frac{\partial^2}{\partial
R^2_*}\right]\phi_k+O(r).
\end{eqnarray}
It shows that the physics near WIH is an infinite collection of
(1+1) dimensional fields in the effective 2-dimensional space-time
with a $U(1)$ gauge field.

Again, the boundary condition of neglecting the ingoing modes near
WIH is imposed. As reviewed in \cite{Be00}, the consistent form of
2-dimensional Abelian anomaly is
\begin{eqnarray}
\del_{\mu}J^{\mu}=\frac{e^2}{4\pi\sqrt{-g}}\
 \eps^{\al\be}\partial_\al A_\be,\qquad \al,\be,\mu=0,1,\label{Gan}
\end{eqnarray}
where $J^{\mu}$ is the current of the $U(1)$ field. The current is
conserved in out region and satisfies above equation in near horizon
region. The general solution of Eq.(\ref{Gan}) for $J^\mu$ is
\begin{eqnarray}
&&J^R_{O}=c_O,\\
&&J^R_H=c_H+\frac{e^2}{4\pi}[A_t(R)-A_t(0)],
\end{eqnarray}
where $c_O$ and $c_H$ are constants. The vanishing variation of the
effective action with respect to the gauge parameter $\Lambda$
\begin{eqnarray}
0&=&-\de W =\int dx^2 \, \Lam\del_\mu J^\mu\nonumber\\
&=&\int dx^2\ \Lam
\left[\de(r-a)\left(J^R_O-J^R_H+\frac{e^2}{4\pi}A_t\right)+\partial_R\left(\frac{e^2}{4\pi}A_t
\Theta_-\right)\right],
\end{eqnarray}
implies
\begin{eqnarray}
c_O=c_H-\frac{e^2}{4\pi}A_t|_{\cH}.
\end{eqnarray}
Further, the covariant current vanishes at the WIH  results in
\begin{eqnarray}
c_O=-\frac{e^2}{2\pi}A_t|_{\cH}=-\frac{e^2}{\pi}\Phi_l . \label{cflux}
\end{eqnarray}
It is just the charge flux.  For Reissner-Nordst\"om black hole, as
an example, $\Phi_l=Q/(2r_+)$ and Eq.(\ref{cflux}) reduces to
Eq.(12) in \cite{IUW06}.  It is worth to point out that above result
is independent on the gauge choice Eq.(\ref{Aga}) because of
Eq.(\ref{Gan}).


\section{Generalization to weakly isolated horizon in higher dimensional spacetime}

The previous discussion can be straightly generalized to the WIH in
a higher dimensional spactime.
\begin{definition}(Weakly Isolated Horizon in Higher Dimensional
Space-time)\\
{An $(n-1)$-dimensional null hypersurface $\cH$ in an
$n$-dimensional spacetime $(M, g)$ is said to be a weakly isolated
horizon (WIH), if\\
1) $\cH$ has the topology of ${\bf K}\times{\bf R}$, where ${\bf K}$
is an
$(n-2)$-dimensional compact Riemannian manifold;\\
2) the expansion of the
null generator of $\cH$, whose tangent vector field is $l^a$, is
zero, i.e. $\Theta_l=0$ on $\cH$;\\
3) $T_{ab}v^b$ is future causal
for any future-directed causal vector $v^a$ and $n$-dimensional
Einstein equations hold in a neighborhood of
$\cH$;\\
4) acting on $l$, Lie derivative $\cL_l$ and induced covariant
derivative $D_a$ on $\cH$ commute, i.e. $[\cL_l\ ,\ D_a]l^b=0$ on
$\cH$.}
\end{definition}
Again, there exists a 1-form $\omega_a$ such that
$D_al^b\heq\omega_al^b$ by definition.

Now, in the Bondi gauge in a neighborhood of $\cH$, $n$-bein can be
expressed in the Bondi-like coordinates $(u,r,\zeta^A)$ with $A, B
=2 , \cdots n-1$ as
\begin{eqnarray}
\begin{cases}
l^a=\partial_u+U\partial_r +X^A\partial_A \cr
n^a=\partial_r\ ,\cr
e^{Aa}=\omega^A \r_r+\xi^{AB} \r_B ,
\end{cases}\label{nbein}
\end{eqnarray}
where
\begin{eqnarray}
U \heq X^A \heq \omega_A \heq 0 . \label{nD-UXom}
\end{eqnarray}
They satisfy the normal condition:
\begin{eqnarray}%
l^a l_a=0, \quad n^a l_a=1, \quad n^a n_a=0,  \quad l^a e_a^A=0,
\quad  n^a e_a^A=0, \quad  e^{Aa} e^B_{\ a}=\de^{AB}.
\end{eqnarray}
The metric of $n$ dimensional spacetime is \cite{NT80}
\begin{eqnarray}
\left(g^{\mu\nu}\right)=\left(\begin{array}{ccc}
0&1&0\\
1 & 2U +\de_{CD}\omega^C \omega^D & X^A + \de_{CD} \omega^C \xi^{DA}\\
0 & X^B + \de_{CD} \omega^C \xi^{DB} & \de_{CD}\xi^{CA}\xi^{DB} \end{array}\right).
\end{eqnarray}
In the Bondi coordinates, $\sqrt{|g|}=\sqrt{h}$ where $(h_{AB}):=(\de_{CD}\xi^{CA}\xi^{DB})^{-1}$.

The Bondi gauge can be expressed as
\begin{eqnarray} \begin{cases}
\nu^A := - n_{\mu;\nu} e^{A\mu} n^\nu = -  e^{A\mu} \Delta n_\mu =0, & \cr
\tau^A :=  l_{\mu;\nu} e^{A\mu} n^\nu = e^{A\mu} \Delta l_\mu =0, & \cr
\kappa^A := l_{\mu;\nu}e^{A\mu} l^\nu = e^{A\mu} D l_\mu \heq 0, & \cr
\gamma  :=  -l^\mu \Delta n_\mu =0, & \cr
\gamma^{[AB]} := e^{[A|\mu|}\De e^{B]}_{\ \mu} = 0, & \cr
\varepsilon^{[AB]} := - e^{[A|\mu|}  D e^{B]}_{\ \mu} \heq 0, 
& \cr
\al^A := n^\mu \de^A l_\mu = - e^{A\mu} D n_\mu =: \pi^A, 
&  \cr
\mu^{[AB]} := n_{\mu;\nu} e^{[A|\mu|} e^{B]\nu} =  e^{[A|\mu|} \de^{B]}n_{\mu}  =0, 
& \end{cases}
\label{nd-gaugec}
\end{eqnarray}
where $D$ and $\Delta$ is defined as before, $\de^A:=e^{Aa} \del_a$.
In the Bondi gauge, 1-form $\omega_a$ is expressed as
\[
\omega_a=-\vps_0 n_a + \de_{AB}\pi^A e^B_{\ a},
\]
where $\varepsilon_0 := n^\mu  D l_\mu \heq \mbox{constant}$ \cite{As04}.
In addition,
the definition of WIH in $n$-dimensional spacetime implies 
\begin{eqnarray}
\rho^{AB} :=  e^{A\mu} \de^{B} l_\mu\heq  0. \label{nd-rhosi}
\end{eqnarray}
\omits{\begin{eqnarray} \begin{cases}
\rho := \de_{AB} e^{A\mu} \de^B l_\mu\heq  0, & \cr
\rho^{[AB]} := - e^{[A|\mu|} \de^{B]} l_\mu\heq  0, & \cr
\sigma^{(AB)} :=  e^{(A|\mu|} \de^{B)} l_\mu -\d \rho {n-2}\de^{AB}\heq  0 . &\end{cases}\label{nd-rhosi}
\end{eqnarray}}
And the angular momentum of WIH in $n$-dimensional spacetime is
\[
J[\varphi]=-\frac{1}{8\pi}\int_{\bf K}(\varphi^a\omega_a)dV_{n-2},
\]
where $\varphi^a$ is a vector field on section ${\bf K}$.  As
before, only $\pi^A$ are related to the angular momentum of WIH.

The commutators of the $n$-bein are
\begin{eqnarray}\begin{cases}
\De D - D\De = \gamma D + \varepsilon \De + \de_{AB}(\tau^A +
\pi^A)\de^B,  & \cr \de^A D-D\de^A = (\al^A -\pi^A)D+\kappa^A \De
+\d \rho {n-2}\de^A + \de_{BC} (\sigma^{(BA)}+ \rho^{[BA]} -
\varepsilon^{[BA]} )\de^C,  & \cr \de^A \De - \De\de^A =  -\nu^A D+
(\tau^A -\al^A) \De +\d \mu {n-2}\de^A +
\de_{BC}(\lambda^{(BA)}+\mu^{[BA]}-\gamma^{[BA]}) \de^C, & \cr \de^A
\de^B -\de^B\de^A =2\mu^{[AB]}D +2\rho^{[AB]} \De
+2\de_{CD}\beta^{C[AB]} \de^D ,&
\end{cases}
\end{eqnarray}
where
\begin{eqnarray}\begin{cases}
\sigma^{(AB)} := \rho^{(AB)} -\d \rho {n-2}\de^{AB} \quad \mbox{with} \quad
\rho := \de_{AB} \rho^{AB}, &\cr
\mu^{AB} := -e^{A\mu} \de^{B} n_\mu, & \cr
\lambda^{(AB)} := \mu^{(AB)} - \d \mu {n-2}\de^{AB} \quad \mbox{with} \quad
\mu := \de_{AB} \mu^{AB}, & \cr
\beta^{ABC} := e^{A\mu}\de^{B} e^{C}_{\ \mu}. &\end{cases}
\end{eqnarray}
They require
\begin{eqnarray}
\begin{array}{llll}
U= \hat \varepsilon_0 r+O(r^2), &  \kappa^A = O(r), &  \beta^{CAB} = O(1), & \lambda^{(AB)} = O(1), \smallskip\\
\varepsilon_0 = \hat \varepsilon_0 + O(r), & \partial_u \xi^{AB} \heq 0, &  \pi^A = O(r), &\omega = O(r). \smallskip \\
\end{array} \label{nd-str}
\end{eqnarray}

The d'Alembert operator can be re-written as
\begin{eqnarray}
\Box&=&g^{ab}\del_a\del_b =(l^an^b+n^al^b+\de_{AB} e^{Aa}e^{Bb})\del_a\del_b\nonumber\\
&=&D\Delta +\Delta D  +(\varepsilon_0{ +}\rho ) \Delta { +} \mu D  { -}\de_{AB}\pi^A\de^B +\Delta_{\bf K} \nonumber\\
&=&2\ppu\ppr+2U\left(\ppr\right)^2+2X^A\ppr\r_A +(\partial_u U)\ppr+(\partial_r X^A)\r_A \nonumber\\
&& +(\vps_0 { +} \rho)\ppr { +} \mu \left (\ppu + U \ppr + X^A \r_A
\right ){ -}\de_{AB}\pi^A \de^B +\Delta_{\bf K}. \label{nd-box}
\end{eqnarray}
where
\[
\De_{\bf K} = \de_{AB}\de^A\de^B - \de_{AB}\de_{CD}\beta^{CAB}\de^D
\]
is the Laplacian on ${\bf K}$. Obviously, it has the same behaviors
as in 4-dimensional spacetime when $r\to 0$. Like the 4-dimension
case, the field $\phi$ can also be expressed into a variable
separated form $\phi(x)=\sum_k\phi_k(u,r)F^{(n-2)}_k$, where
$\{F^{(n-2)}_k\}$ are eigenfunctions of the $(n-2)$-dimensional
Laplacian. Introducing tortoise-like coordinates $(t,R)$ as
(\ref{tR}), the previous discussions are still valid for the WIH in
the higher dimensional spacetime.


\section{Conclusion and discussion}
In above work, have shown that the anomaly analysis of Wilczek and
his collaborators is indeed applicable to the general weakly isolated
horizons. Here, we want to give some remarks on our work.

First, we focus on 4-dim spacetime without cosmological constant in
previous work. What we want to emphasis is our derivation also holds
for the spacetime with a nonzero cosmological constant. The reason
is as follows: Suppose we have a 3-dim null surface $\cH$ in a
spacetime with nonzero cosmological constant, such that $\cH$
satisfies all requirements of definition 1. The Einstein equation
with a cosmological constant is
\begin{eqnarray}
G_{ab}-\Lambda g_{ab}=8\pi T_{ab}.
\end{eqnarray}
We can always move the cosmological constant term to the right hand
side of the equation and define a new energy-momentum tensor
$\tilde{T}_{ab}:=T_{ab}+\Lambda g_{ab}/(8\pi)$. From the definition
of WIH, we have known that $T_{ab}$ satisfies $T_{ll}=0$,
$T_{lm}=T_{l\bm}=0$ (or $T_{lA}=T_{Al}=0$ in higher dimensional
spacetime). It is easy to check $\Lambda g_{ab}$ also satisfies
these requirements, so the total energy-momentum tensor
$\tilde{T}_{ab}$ satisfies the requirements of the definition of
WIH. Furthermore, when a positive cosmological constant is present,
the spacetime will, in general, have a cosmological horizon, which
also satisfies the weakly isolated condition. Therefore, the above
derivation also applies to the cosmological horizon. The only
difference is that the topology of WIH for negative cosmological
constant case is $S\times {\bf R}$, where $S$ can be any kinds of
2-dim compact manifold\cite{As04}.

Second, it is worth to notice that our calculation is just based
on Cartan structure equations. This means our calculation can also
be used to black holes from other gravitational theories if the
black hole horizon is null and has zero expansion, for example the
black ring solution, $f(R)$ theory
and other theories. This means the Hawking radiation is in fact a
kinematical effect of spacetime \cite{unruh1,vi1,vi2,cg,Sun}.

\section*{Acknowledgments}
We would like to thank Prof. Z. Chang and Prof. H.-Y. Guo for
helpful discussion. The project is partly supported by the Natural
Science Foundation of China under Grant Nos. 90403023, 10605006,
10705048, 10775140, 10731080 and Knowledge Innovation Funds of CAS
(KJCX3-SYW-S03).

\end{document}